\documentclass[twocolumn,showpacs,preprintnumbers,amsmath,amssymb]{revtex4}
\usepackage{graphicx}% Include figure files
\usepackage{dcolumn}% Align table columns on decimal point
\usepackage{bm}% bold math

\begin{document}

%\preprint{APS/123-QED}

\title{The Decay of Pure Quantum Turbulence in Superfluid $^3$He-B}

\author{D.I.~Bradley}
\author{D.O.~Clubb}
\author{S.N.~Fisher}
\email[Electronic Address: ]{s.fisher@lancaster.ac.uk}
\author{A.M.~Gu\'enault}
\author{R.P.~Haley}
\author{C.J.~Matthews}
\author{G.R.~Pickett}
\author{V.~Tsepelin}
\author{K.~Zaki}
\affiliation{Department of Physics, Lancaster University,
Lancaster, LA1 4YB, UK. }

\date{\today}

\begin{abstract}
We describe measurements of the decay of pure superfluid turbulence in superfluid $^3$He-B, in the
low temperature regime where the normal fluid density is negligible. We follow the decay of the
turbulence generated by a vibrating grid as detected by vibrating wire resonators.  Despite the
absence of any classical normal fluid dissipation processes, the decay is consistent with
turbulence having the classical Kolmogorov energy spectrum and is remarkably similar to that
measured in superfluid $^4$He at relatively high temperatures. Further, our results strongly
suggest that the decay is governed by the superfluid circulation quantum rather than kinematic
viscosity.

\end{abstract}

\pacs{67.57.Fg, 67.57.De, 67.57.Hi}

\maketitle

In this paper we present the first quantitative measurements of the decay of turbulence in a pure
superfluid system. This is a subject of considerable interest since no conventional dissipation
mechanisms are available.

In a classical fluid, turbulence at high Reynolds numbers is characterized by a range of eddy sizes
obeying the well-known Kolmogorov spectrum. On large length scales the motion is dissipationless,
whereas on small scales viscosity comes into play. Decay of the turbulence proceeds as energy is
transferred by non-linear interactions from the largest non-dissipative length scales $d$
(typically the size of the turbulent region) to smaller length scales where the motion is
dissipated by viscous forces. The dissipation per unit volume is given by $\rho \nu \omega^2$ where
$\rho$ is the fluid density, $\nu$ the kinematic viscosity and $\omega^2$ the mean square vorticity
\cite{Vinen}. An interesting question, which has received much theoretical speculation\cite{Vinen},
is what happens in a pure superfluid with no viscous interactions?

Conceptually, turbulence in a superfluid is greatly simplified. Superfluids such as He-II and
$^3$He-B are described by macroscopic wavefunctions with a well defined phase $\phi$. The
superfluid velocity is determined by gradients of the phase, $v_S=(\hbar/m)\nabla \phi$ where $m$
is the mass of the entities constituting the superfluid (the mass of a $^4$He atom for He-II or
twice the mass of a $^3$He atom, $2m_3$, for the Cooper pairs in $^3$He-B). Consequently, in
contrast to classical fluids, superfluid motion is inherently irrotational and vorticity may only
be created in the superfluid by the injection of vortex lines. A superfluid vortex is a line defect
around which the phase changes by $2 \pi$ (ignoring here more complex structures such as in
$^3$He-A). The superfluid order parameter is distorted within the relatively narrow core of the
vortex where all the circulation is concentrated. The superfluid flows around the core with a
velocity, at distance $r$, given by $v_S=\hbar/mr$ corresponding to a quantized circulation $\kappa
= h/m$. Vortex lines are topological defects. They cannot terminate in free space, and therefore
must either form loops or terminate on container walls. Turbulence in a superfluid takes the form
of a tangle of vortex lines.

Superfluid hydrodynamics is further simplified by the superfluid component having zero viscosity.
At finite temperatures the fluid behaves as a mixture of two fluids, the superfluid condensate
component as discussed above and an interpenetrating normal fluid comprising the thermal
excitations. The normal fluid component has a finite viscosity and exerts a damping force on the
motion of vortex lines via the scattering of thermal excitations, this interaction being known as
mutual friction.

To date, studies of superfluid turbulence have largely focussed on He-II at relatively high
temperatures. Under these conditions, it is believed that mutual friction effectively couples the
turbulent structures in the normal and superfluid components \cite{Vinen}. The ensuing combined
turbulence is found to behave in an almost identical manner to that of classical turbulence when
generated by a towed grid \cite{stalp,skrbek}.  The decay of grid turbulence observed in He-II can
thus be explained quantitatively \cite{stalp,skrbek} using the classical picture with the
conceptually reasonable assumptions that $\omega^2=(\kappa L)^2$ where $L$ is the length of vortex
line per unit volume, and that the effective kinematic viscosity is $\nu \sim \eta_n/\rho$ where
$\eta_n$ is the normal fluid viscosity and $\rho$ is the total fluid density.

The situation in superfluid $^3$He should be completely different. The fermionic nature of normal
liquid $^3$He ensures that the liquid is very viscous (comparable to room temperature glycerol).
This high normal fluid viscosity means that the normal component can never become turbulent under
typical experimental conditions. Further, owing to the interaction via mutual friction, turbulence
in the superfluid is also suppressed at high temperatures. Consequently, turbulence in $^3$He-B is
only found at temperatures below $\sim 0.5T_c$ where the mutual friction has become low enough to
decouple the two components, allowing the superfluid to support turbulence independently
\cite{Helsinki}.

At even lower temperatures (below $\sim 0.3 T_c$) both the normal fluid component and mutual
friction become exponentially small, the excitations are too dilute to interact and become
ballistic. In this regime, the whole concept of a normal fluid component breaks down. These are
conceptually the simplest conditions for studying turbulence; we effectively have only one
incompressible and irrotational fluid component with zero viscosity supporting quantized vortex
lines. Here we have a system where the classical decay mechanism {\it absolutely} cannot operate.
So, what happens instead?

Turbulence in superfluid $^3$He-B can be readily detected at low temperatures via its effect on the
quasiparticle dynamics \cite{hale}. The dispersion curve $\epsilon({\bf p})$ of these ballistic
quasiparticles is tied to the reference frame of the stationary superfluid. The curve thus becomes
tilted by the Galilean transformation $\epsilon({\bf p}) \rightarrow \epsilon({\bf p})+ {\bf
p}\cdot {\bf v_S}$ in a superfluid moving with velocity ${\bf v_S}$. Consequently, quasiparticles
moving along a superflow gradient experience a potential energy barrier and are Andreev reflected
if they have insufficient energy to proceed \cite{andreev}. The Andreev process converts a
quasiparticle into a quasihole and vice versa, reversing the group velocity of the excitation but
yielding negligible momentum transfer.

The complicated flow field associated with superfluid turbulence acts as a shifting ragged
potential for quasiparticles.  The net result is that some fraction of incident thermal
quasiparticles are Andreev reflected. Quasiparticles may be detected in $^3$He-B at low
temperatures by vibrating wire techniques. The thermal damping of a vibrating wire \cite{carney} in
$^3$He-B arises from normal scattering of quasiparticle excitations at the wire surface. A wire
immersed in turbulence thus experiences a reduction in damping proportional to the amount of
Andreev reflection of incoming thermal excitations caused by the turbulent flow. This effect has
been exploited to observe turbulence generated by vibrating wires \cite{hale} and vibrating grid
\cite{gridrings} resonators at low temperatures. Andreev reflection from turbulence has also been
measured directly using ballistic quasiparticle beam techniques \cite{bradley}. Previous
measurements of vortex generation by a vibrating grid have shown that at low grid velocities
ballistic vortex rings are emitted \cite{gridrings} and turbulence only forms above a certain
critical velocity. Here, we discuss measurements of the decay of turbulence generated from a
vibrating grid at higher velocities.

The experimental arrangement is shown in figure 1 and is the same as that used for the measurements
reported previously \cite{gridrings, gridresponse}. The grid is made from a 5.1$\times$2.8\,mm mesh
of fine copper wires. The wires have an approximately 11\,$\mu$m square cross-section and are
spaced 50\,$\mu$m apart leaving 40\,$\mu$m square holes. A 125\,$\mu$m diameter Ta wire is bent
into a 5\,mm square frame and attached to the inner cell wall of a Lancaster style nuclear cooling
stage \cite{stage}. The mesh is glued to the Ta wire over thin strips of cigarette paper for
electrical insulation.

Facing the grid are two vibrating wire resonators made from 2.5\,mm diameter loops of 4.5\,$\mu$m
NbTi wire. The `near' and `far' wires are positioned 1\,mm and 2\,mm from the grid respectively. An
additional wire resonator is used as a background thermometer. This wire, not shown in the figure,
is located about 4\,mm to the side of the grid and enclosed in a mesh cage to ensure that its
response is not influenced by any stray turbulence.

The grid is operated similarly to a wire resonator. It is situated in a vertical applied magnetic
field and driven by the Lorentz force generated by passing an ac current through the Ta wire. As
the grid moves, the Ta wire develops a Faraday voltage proportional to its velocity. The grid
resonates at a frequency of $\sim$1300\,Hz, determined by the stiffness of the Ta wire and the mass
of the grid.

\begin{figure}
\includegraphics[width=0.7\linewidth]{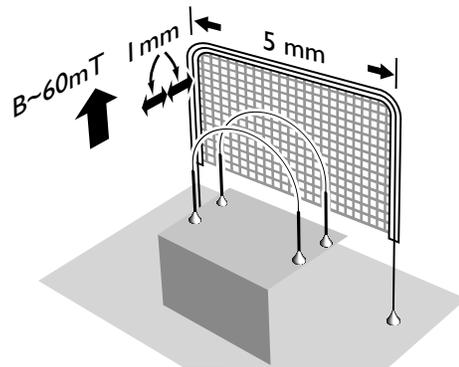} \caption{The arrangement of the grid and associated vorticity detector wires.}
\end{figure}

In contrast to a vibrating wire resonator, the grid shows no sign of a pair-breaking critical
velocity. In the low temperature limit, the grid's response changes gradually from a linear damping
force $F\propto v$ for velocities below around 1\,mm/s, to approximately $F\propto v^2$ behavior at
higher velocities \cite{gridresponse}. The linear response is governed by the intrinsic (vacuum)
damping of the resonator motion. The response at high velocities has the form expected for
turbulent drag from a classical fluid \cite{gridresponse}.

Vortices generated by the grid are detected by the two facing vibrating wire resonators as
discussed in \cite{gridrings}. Briefly, the two resonators and the thermometer resonator are driven
on resonance at relatively low velocity. The resulting induced voltages across the wires are
continuously monitored, allowing us to deduce the quasiparticle damping (frequency width of the
resonance) $\Delta f_2(T)$ for all three wires. The grid is then driven to some velocity $v$
generating vortex lines (ballistic vortex rings at low velocities; turbulence at higher
velocities). This vorticity Andreev-reflects some fraction $f$ of quasiparticles approaching a
vibrating wire, giving rise to a reduced damping $\Delta f_2(v,T)=(1-f)\Delta f_2(0,T)$. In
practice, significant power is required to drive the grid, resulting in an overall warming of the
cell.  The damping in the absence of turbulence $\Delta f_2(0,T)$ is therefore inferred from the
thermometer wire damping (with no turbulence, the quasiparticle damping on each of the three wires
is simply related by a measured constant of proportionality, close to unity). The fractional
screening $f$ of quasiparticles due to the surrounding turbulence is thus measured for the two
facing wires.

All the measurements discussed below were made at 12\,bar and at temperatures below $\sim
0.2\,T_c$. At such temperatures the turbulence is found to be insensitive to temperature. This is
consistent with previous measurements, both of turbulence generated from vibrating wires
\cite{bradley} and of vortex rings generated from a vibrating grid \cite{gridrings}, indicating
that we have reached the zero temperature limit for the turbulent dynamics where both the normal
fluid fraction and the mutual friction are negligible.

The steady state average values of the fractional screening $f$ are found to increase roughly as
$v^2$. The `far' wire, 2\,mm from the grid, has roughly a factor of two less screening than the
`near' wire, 1\,mm from the grid, over the entire velocity range. If the variation with distance
followed an exponential decay, as found previously for turbulence generated by vibrating wires
\cite{spatial}, then this would correspond to a spatial decay length of $d\sim1.5$\,mm.

The approximate vortex line density may be inferred from these measurements using the arguments of
\cite{bradley}. The fraction of quasiparticles Andreev reflected after passing through a homogenous
isotropic vortex tangle of line density $L$ and thickness $x$ is given by $f\simeq L p_F \hbar x /
2 m_3 k_B T$ provided $f$ is small compared to unity. Since in practice the tangle density varies
in space, strictly we should integrate an analogous expression over all quasiparticle trajectories
incident on the vibrating wire resonators. This is obviously not possible without an accurate
knowledge of the spatial dependence of the tangle. We therefore simply use the above expression
with $x=d=1.5$\,mm to give an estimated average line density which should be correct to within a
factor of order 2.

The transient behavior of the inferred line density after the drive to the grid is turned off is
shown in figure 2 for the wire nearest to the grid. Data are shown for various initial grid
velocities down to 3.5\,mm/s. (At lower velocities the recovery is much faster corresponding to
ballistic vortex ring production \cite{gridrings}.) At late times the data all tend a single
limiting line (line A in the figure).

\begin{figure}
\includegraphics[width=0.8\linewidth]{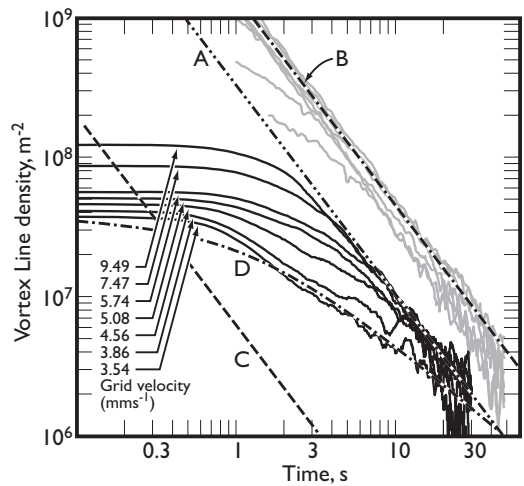}
\caption{\label{fig:BCScurves} Solid Black curves show the inferred vortex line density as a
function of time after cessation of grid motion for various initial grid velocities. Line A is the
limiting behavior scaled to our results as discussed in the text. The halftone data is that for a
towed grid in superfluid $^4$He of Skrbek {\it et. al} \cite{skrbek} with line B showing the
late-time limiting behavior. Line C shows the expected behavior for our data assuming the classical
dissipation law. Curve D shows the expected behavior for a random tangle in superfluid $^3$He. See
text.}
\end{figure}

In figure 2 we also show data for turbulent decay from a grid towed at various velocities through
He-II \cite{skrbek}. The authors shifted the time axis for each of these curves, but this does not
effect the late time behavior which is fitted by line B \cite{stalp, skrbek} (see below). The
fitted line lies about a factor of 4 higher than our data. The authors were able to explain these
observations in some detail on the basis of classical turbulence of the combined normal/superfluid
components. The classical cascade process leads to a line density which decays as $L=(d/2\pi
\kappa) \sqrt{(27C^3/\nu^\prime)} \, t^{-3/2}$ at late times \cite{stalp, skrbek}, where $C$ is the
Kolmogorov constant, expected to be of order unity and $d$ is the characteristic size of the
container (which limits the maximum eddy size in the classical theory). Excellent agreement was
found with their data using $C\simeq1.6$ and an effective kinematic viscosity $\nu^\prime$ of
roughly twice the actual kinematic viscosity, $\eta_n/\rho$.

If we take a similar approach and naively use this classical expression for the late-time line
density, substituting the appropriate numbers for our experiment then we obtain line C in Fig. 2.
This line lies much lower than that of $^4$He partly since dimension $d$ is smaller ($d=$1.5\,mm in
our case against $d$=10\,mm for the $^4$He experiments) but mainly because the normal fluid
viscosity \cite{CHH} is orders of magnitude larger for $^3$He. It is very clear that our
measurements, even though they are similar to those in superfluid $^4$He, {\it cannot} be explained
by the classical decay mechanism, as we anticipated.

The Kolmogorov energy cascade in classical turbulence is a consequence of dissipation being
negligible on large length scales. As suggested by Vinen \cite{Vinen}, it seems reasonable to
expect that superfluid turbulence as generated by grids will display a similar cascade process
owing to the similar absence of large length scale dissipation mechanisms. This expectation is
supported by numerical simulations \cite{TsubotaKolmogorov} which show evidence of a
Kolmogorov-like cascade in pure superfluid turbulence in the absence of any normal-fluid component.
In other words, for He-II both fluid components have a natural tendency to display the
Kolmogorov-like cascade. Therefore this behavior is likely to occur at arbitrary temperatures, and
with the two flows locked together by mutual friction at the higher temperatures. By the same
reasoning, one might expect similar behavior for superfluid $^3$He-B in the low temperature limit.
At the higher temperatures, mutual friction will now couple the superfluid turbulence to the highly
viscous non-turbulent normal $^3$He, suppressing turbulence completely at high temperatures, and
yielding a different energy spectrum in the intermediate region \cite{Volovik}.

At very low temperatures in the superfluid where there are no mutual friction processes, Vinen
\cite{Vinen} has argued (on purely dimensional grounds) that any process leading to loss of
vortex-line length must depend on the circulation quantum, yielding a dissipation of order $\rho
\kappa (\kappa L)^2$. The effective kinematic viscosity in the decay equation should therefore be
replaced by a term $\zeta \kappa$ where $\zeta$ is a dimensionless constant, presumably of order
unity. The line density at late times of the turbulent decay should therefore be described by
$L=(d/2\pi \kappa) \sqrt{(27C^3/\zeta \kappa)} \, t^{-3/2}$.

Since in He-II the kinematic viscosity and the circulation quantum are numerically similar
($\nu\approx0.1\kappa$), the data of Skrbek {\it et al} \cite{skrbek} interpreted above on the
basis of the kinematic viscosity are also consistent with a dissipation based on the quantum
expression with $\zeta \approx 0.2$. However, $\nu$ and $\kappa$ are orders of magnitude different
in superfluid $^3$He. If we use the Vinen expression for our data, with $d=1.5$\,mm and
$\zeta=0.2$, then we obtain the expected late-time behavior shown by line A in the figure.
(equivalent to scaling the late-time He-II data by $d$ and $\kappa$). The agreement is quite
staggering, since not only does the superfluidity in the two systems arise from completely
different mechanisms, but both the temperature regimes and normal fluid viscosities differ by many
orders of magnitude.

The decay for the lowest grid velocity shown in Fig. 2 appears to show a limiting behavior closer
to $t^{-1}$. A purely random tangle can have only one length scale, that of the intervortex spacing
$L^{-1/2}$ and hence no Kolmogorov cascade. In this case the line density is expected to decay by
the Vinen equation \cite{Vinen} $\dot{L}=\zeta^\prime \kappa L^2$. Curve D in the figure shows the
expected behavior according to this equation with $\zeta^\prime=0.3$ and an initial line density
chosen to match the lowest grid velocity data at the start of the decay. The agreement is fair,
suggesting that the Kolmogorov energy cascade might only develop for higher grid velocities (line
densities). This is not conclusive however, since the lower grid velocity data could also be made
to fit with the full classical model given in \cite{skrbek}.

As a final caveat, if the turbulence we generate is inhomogeneous then the observed
decay may include a spatial component from the diffusion of the vorticity down a vorticity
gradient. However, we can estimate this effect from the computer simulations by Tsubota {\it et.
al.} \cite{tsubota} which suggest that inhomogeneous turbulence evolves spatially with a diffusion
constant of $\sim 0.1\kappa$. For our experiment this number yields a time scale for diffusion of
order $\tau \sim d^2/0.1\kappa \sim 300$s. This is much longer than the measured decay time and
therefore any contribution from diffusion should not be significant. (We also note that turbulence
generated in classical fluids by oscillating grids can be quite isotropic under certain conditions
\cite{classicalgrids}.)

In conclusion, we have measured the decay of turbulence in superfluid $^3$He-B generated by a
vibrating grid at very low temperatures where there is essentially no normal fluid. The decay is
found to be consistent with a classical Kolmogorov-type energy cascade and very similar to that
found for turbulence from a towed grid in He-II at high temperatures. This is a remarkable result
given that the two liquids have entirely different mechanisms for superfluidity and that the
measurements were performed at opposite ends of the temperature range. In contrast to the He-II
case, the decay observed in these measurements {\it cannot} be explained in terms of a classical
decay mechanism (i.e. via a normal fluid viscosity).  The measurements strongly indicate that the
decay is governed by the circulation quantum, which has a similar magnitude to that of He-II. The
questions remaining are: a) what is the specific microscopic mechanism for the dissipation and b)
how does the superfluid tangle acquire or develop the requisite range of length scales necessary
for the Kolmogorov energy cascade to function?

We acknowledge financial support from the UK EPSRC, excellent
technical support from I.E.~Miller and M.G.~Ward, and useful
discussions with C.~Barenghi, L.Skrbek, M.Tsubota and W.F.Vinen.

\end{document}